\documentclass[12pt]{article}

\usepackage{graphicx} 
\usepackage{jheppub}
\usepackage{amsfonts,amsmath} 
\usepackage{mathrsfs} 
\usepackage{setspace}
\usepackage[utf8]{inputenc} 
\usepackage{ytableau}
\usepackage{comment} 

\makeatletter
\def\@fpheader{\relax}
\makeatother

\DeclareMathAlphabet{\mathbbold}{U}{bbold}{m}{n} 
\newcommand{\be}{\begin{equation}} \newcommand{\ee}{\end{equation}}

\newcommand{\thalf}{{\tfrac{1}{2}}}
\newcommand{\lie}[1]{{\mathsterling}_{\raisebox{-1pt}{\!$\scriptstyle #1$\,}}}

\DeclareMathOperator{\Tr}{Tr}

\newcommand{\half}{\frac{1}{2}}

\title{Large gravitons and near-horizon diffeomorphisms} 
\author{Bruno Carneiro da Cunha,}\emailAdd{bcunha@df.ufpe.br}
\author{Filipe Rudrigues}\emailAdd{frudrigues@gmail.com}

\affiliation[a]{Departamento de Física, Universidade Federal de Pernambuco,
50670-901, Recife, Pernambuco, Brazil} 

\abstract{Usual gauge fixing procedures in classical general
  relativity rely on the existence of solutions of a second order wave
  equation. We propose to use these equations to relate asymptotic
  symmetries at infinity to asymptotic symmetries of a black hole
  horizon, in tune with recent proposals. We illustrate the
  construction for the BTZ and four-dimensional Kerr black holes. We
  find in both cases a realization of the group of diffeomorphisms of
  the real line.
}

\preprint{\today}

\begin{document}

\maketitle

\section{Introduction}
\label{sec:introduction}

Black holes are seemingly simple objects: most of known examples
involve exact solutions for Einstein's equations, which is something
rare. In four dimensions and lower, uniqueness theorems guarantee that
black hole metrics are the natural endpoint for the evolution of
gravitational systems, whereas entropic arguments posit the same fate
even in higher dimensions. No-hair theorems corroborate this point of
view by stating that the classical black hole is completely
characterized by its observable charges.

There is, however, a major difference, between a black hole and this
``structureless particle'' point of view: the existence of an event
horizon. This furnishes the black hole with a structure which can be
seen both as geometric and thermodynamic, with profound implications
for its physical description. The event horizon is a surface of infinite
redshift, and as such its description defies the ``effective field
theory'' point of view pervasive from other branches of high energy
physics. 

The black-hole horizon is usually defined classically as a global
property of the space-time: the boundary of the causal future of the
``asymptotic past'' and the causal past of the ``asymptotic future''. 
Although we should point out at this point that alternative
definitions do exist: particularly the notion of trapped surface,
which has the advantage of being local, and in fact is much more
manageable, specially in numerical simulations. The terms ``asymptotic
past'' and ``asymptotic future'' deserve further qualification: they
fall into a more generic notion of ``asymptotic infinity'', which
tries to define the analogue of an ``isolated system'' when gravity is
present \cite{Geroch:1977jn}.  

The notion of asymptotic infinity accomplishes this analogue, at the
expense of giving up some of the diffeomorphism invariance. The
analogue, as well as the fixing of the diffeormorphism invariance,
seem necessary for a proper definition of space-time observable
charges, like energy or angular momentum, which depends 
explictly on a notion of Poincaré invariance. In turn, Poincaré
invariance is expected to hold only infinitely far from the sources,
so, in dealing with non-trivial cases, one arrives at the notion of
``asymptotic isometry'' \cite{Sachs:1962zza}. The most
famous examples are the Bondi-Metzner-Sachs (BMS) group in four
dimensional general relativity with zero cosmological constant
\cite{Bondi:1962px,Sachs:1962wk} and the Brown-Henneaux (BH) group in
three-dimensional  gravity with negative cosmological constant
\cite{Brown:1986nw}. In both cases, one cannot extricate the ``global
isometries'' -- the isometries of the vacuum solutions -- from the
group of asymptotic symmetries. In the Brown-Henneaux case, 
the charges associated to the asympotic isometries generate a
non-trivial Virasoro algebra, which gives a consistent geometric
interpretation for the physical degrees of freedom associated with the 
black hole background.

In terms of local physics, this attribution shares many parallels with
the situation in non-abelian gauge theories, where the local degrees
of freedom are often complemented by ``large gauge transformations''.
Would-be gauge transformations -- in the gravity case, coordinates
transformations -- which are no longer duplicate descriptions of the same
physical configuration since they change the value of observables of
the system. Unlike instantons in non-abelian gauge fields, however,
there is no gauge-invariant way to think about those transformations
as localized, nor any known topological invariant associated to
it. Their association with {\it bona-fide}, physical local degrees of
freedom of the black hole is then problematic, even when the numerical
checks seem to match, as in the Brown-Henneaux case.  

Proposals for local, geometric diffeomorphisms that count the black
hole degrees of freedom (``black hole hair'') along the directions
above have been put forward, by a number of authors over the
years. See \cite{Strominger:1997eq,Carlip:1999cy} for examples
relevant to our discussion. They have been, however, always plagued
by the unclear message of gauge invariance. Recently Hawking and
collaborators proposed to tackle the problem by transposing the
concept of asymptotic isometry to the horizon
\cite{Hawking:2016msc,Hawking:2016sgy} -- see also Donnay {\it 
  et al.} \cite{Donnay:2015abr,Donnay:2016ejv} -- which could in
principle solve not only the problem of counting the
Bekenstein-Hawking entropy formula but also the information paradox 
problem.  

In this article we propose that the problem of gauge invariance can be
consistently solved by linking the approximate isometries at infinity
to the ones at the horizon, via the gauge fixing conditions. This
recipe, though dependent on the particular dynamical model whose
solution is the black hole under consideration, has the advantage of
expliciting the gauge choices involved in ascribing the
diffeomorphisms, while keeping the local aspect of the proposal. We
will revise some of the important notions in Sections 2 and 3, and
work out explictly the BTZ and Kerr cases in Section 4 and 5,
respectively. We close with a summary and some prospects.

\section{Asymptotics and symmetries}

Many papers and textbooks cover the issues of asymptotic simplicity,
asymptotic isometries and supertranslations, see, for instance,
\cite{Geroch:1977jn} and chapter 11 in \cite{Wald:1984}. We will focus
on the anti-de Sitter and flat case. In both of them, there is a
non-vanishing $\Omega$ function which serves to link the physical
metric $g_{ab}$ to a non-physical metric $\hat{g}_{ab}$ by means of a
conformal transformation:
\begin{equation}
g_{ab}=\Omega^{-2}\hat{g}_{ab},
\end{equation}
in such a way that the asymptotic region -- far from sources -- can be
mapped to the pre-image of $0<\Omega<\epsilon$. The ``conformal
boundary'' $\Omega\rightarrow 0$ is a region added to the unphysical
manifold in such a way that it is topologically closed. The causal
structure of the conformal boundary depends on the model considered:
if the cosmological constant is negative, the boundary is space-like
-- save for two points -- with the topology of a cylinder. If the
cosmological constant is zero, the boundary is divided in five pieces,
two topologically given by $\mathbb{R}\times S^2$, called past and
future null infinity, and three points, past and future timelike
infinity and spacelike infinity. Obviously many issues about the
asymptotic behavior of fields are ``swept under the rug'',
particularly the behavior at the ``boundaries of the boundaries'':
both timelike infinites, and spacelike infinity in the case of
asymptotically flat spacetimes; time-like infinities in the case of
negative cosmological constant. With all its shortcomings, the
procedure is rich enough to tackle the issues raised in the
introduction.

From the definition of the conformal boundary as the $\Omega=0$ level
surface a lot of structure arises.  Consider vacuum Einstein's
equation written in terms of the unphysical metric $\hat{g}_{ab}$:
\begin{gather}
R_{ac}=\frac{k(D-1)}{\ell^2}g_{ac}, \quad\text{or} \\
\hat{R}_{ac}+
(D-2)\frac{\hat{\nabla}_a\hat{\nabla}_c\Omega}{\Omega} 
-(D-1)\hat{g}_{ac}\hat{g}^{bd} 
\frac{\hat{\nabla}_b\Omega\hat{\nabla}_d\Omega}{\Omega^2}+
\hat{g}_{ac}\hat{g}^{bd}\frac{\hat{\nabla}_b\hat{\nabla}_d\Omega}{\Omega}
=\frac{k(D-1)}{\ell^2}\frac{\hat{g}_{ac}}{\Omega^2},
\label{eq:einsteinconformal}
\end{gather}
where in the second line we wrote the Ricci tensor associated with
$g_{ac}$ in terms of hatted quantities, which are associated with
$\hat{g}_{ac}$. In the first equation $k=0,+1,-1$ corresponding to the
flat \cite{Wald:1984},  de Sitter and anti-de Sitter
\cite{Ashtekar:1984zz} cases respectively. Regularity of
the unphysical metric at $\Omega=0$ requires that, at the boundary: 
\begin{equation}
\hat{g}^{bd}\hat{\nabla}_b\Omega\hat{\nabla}_d\Omega=
-\frac{k}{\ell^2}\quad \text{at}\quad\Omega=0,
\end{equation}
so the normal vector $n^a=\hat{g}^{ab}\hat{\nabla}_b\Omega$ to the
$\Omega=0$ surface will be spacelike for negative cosmological
constant, timelike for positive cosmological constant and null for the
flat case. By multiplying $\Omega$ by a non-vanishing function
$\omega$ at the boundary one can further ensure that, near $\Omega=0$,
\begin{equation}
\hat{g}^{bd}\hat{\nabla}_b\Omega\hat{\nabla}_d\Omega=-\frac{k}{\ell^2}+{\cal
  O}(\Omega^2),
\end{equation}
and therefore the vanishing of the ${\cal O}(\Omega^{-1})$ term in
\eqref{eq:einsteinconformal} requires:
\begin{equation}
\hat{\nabla}_a\hat{\nabla}_c\Omega=0\quad\text{at} \quad\Omega=0,
\end{equation}
which in turn implies that the vector $n^a$ is covariantly constant at
$\Omega=0$. This construction still allows for a gauge symmetry: one
may still multiply the $\Omega$ function by a nowhere vanishing
function $\omega$: $\Omega\rightarrow \omega\Omega$, which is constant
along $n^a$: $n^a\hat{\nabla}_a\omega=0$. This remaining conformal
symmetry allows us to fix the conformal structure of the boundary, 
and, by choosing a particular set of coordinates (``Bondi
coordinates'' \cite{Tamburino:1966zz}), write a metric for the
asymptotic boundary as induced by the interior non-physical metric
$\hat{g}_{ab}$. This induced metric can be fixed to be that of a flat
cylinder for asymptotic anti-de Sitter spaces $k=-1$ and a null line
times a standard, constant curvature sphere for the asymptotically
flat case $k=0$. More importantly, the remaining conformal symmetry of
the unphysical metric gives rise to the conformal symmetries. The
usual isometry condition $\lie{\xi}g_{ab}=0$ can only be expected to
hold at $\Omega=0$ (``infinitely far away''), so we define an
asymptotic symmetry as a vector field $\xi^a$ which has a smooth limit
to the surface $\Omega=0$ such that 
\begin{equation}
\Omega^2\lie{\xi}g_{ab}=0, \quad\text{at}\quad\Omega=0.
\label{eq:confisos}
\end{equation}
Expanding the Lie derivative in terms of the unphysical metric
we find \cite{Frolov:1979ab}: 
\begin{equation}
\hat{\nabla}^a\xi^b+\hat{\nabla}^b\xi^a-2\hat{g}^{ab}
\frac{\xi^c\hat{\nabla}_c\Omega}{\Omega}=0
\quad\text{at}\quad\Omega=0, 
\end{equation}
which can be used to recover the asymptotic symmetries of vacuum space
-- the (anti)-de Sitter group for $k\neq 0$ and the Poincaré group for
$k=0$. Because the condition is only enforced at $\Omega=0$, we have
an equivalence class of solutions: two solutions of
\eqref{eq:confisos} $\xi^a$ and $\xi'^a$ generate the same asymptotic
symmetry if the vector fields coincide at $\Omega=0$.

Now we can specialize to the two cases we are going to address here:

\subsection{Three-dimensional anti-de Sitter space-times}

In this case one can find coordinates $z,u,v$ such that the metric has
the asymptotic form:
\begin{equation}
ds^2=\frac{dz^2+du\,dv}{z^2},
\end{equation}
where one takes $\Omega=z$ and then the unphysical metric
$\hat{g}_{ab}$ is locally three-dimensional Minkowski. The generic
solution of \eqref{eq:confisos} is given by the Brown-Henneaux
generators \cite{Brown:1986nw}:
\begin{equation}
\ell_n=\frac{1}{2}e^{2nu}\left(-n\,z\frac{\partial}{\partial
    z}+\frac{\partial}{\partial u}\right) ,\quad\quad
\bar{\ell}_n=\frac{1}{2}e^{2nv}\left(n\,z\frac{\partial}{\partial
    z}+\frac{\partial}{\partial v}\right) ,\quad n\in\mathbb{Z},
\label{eq:brownhenneaux}
\end{equation}
with each set satisfying the Witt algebra:
\begin{equation}
[\ell_n,\ell_m]=-(n-m)\ell_{n+m}.
\end{equation}
The generators $\ell_{-1},\ell_0,\ell_1$ and their barred counterparts
generate the three-dimensional anti-de Sitter algebra ${\rm
  SL(2,\mathbb{R})\times SL(2,\mathbb{R})}$. The remaining operators
induce generic, local conformal transformations at spatial
infinity $i^0$. Heuristically, scale transformations of $i^0$ induced
by $\ell_n$, $\bar{\ell}_n$ can be ``undone'' by a translation of
$z$ \cite{Ashtekar:1984zz}.

\subsection{Four-dimensional asymptotically flat space-times} 

Now $n^a$, the normal vector to the surface $\Omega=0$ is null, and
one can construct coordinates $u,\Omega,\theta,\phi$ such that the
unphysical metric at future null infinity is given by:
\begin{equation}
d\hat{s}^2=2du\,d\Omega+d\theta^2+\sin^2\theta\,d\phi^2.
\end{equation}
In these coordinates $n^a=\partial/\partial u$. Along with the usual
Poincaré symmetries, one can check that $\xi^a=\alpha n^a$ satisfies
that $\Omega^2\lie{\xi}g_{ab}=0$ at $\Omega=0$ for generic $\alpha$
satisfying $n^a\hat{\nabla}_a\alpha=\partial_u\alpha=0$. Since
$\alpha$ is now a function of $\theta$ and $\phi$ only it can be
expanded in spherical harmonics. The $l=0$ and $l=1$ pieces completes
the Poincaré group whereas the higher harmonics form an abelian
algebra called ``supertranslations''. The whole space of solutions is
called the BMS group. 

One important fact about supertranslations for the following analysis
is that the solutions $\xi^a=\alpha n^a$ can be obtained from a
potential. We note that the gradient of $\Phi=\alpha\Omega$ with
$\alpha$ as above induces a vector field:
\begin{equation}
\xi^a=\hat{g}^{ab}\hat{\nabla}_b\Phi=\alpha n^a+\Omega
\hat{g}^{ab}\hat{\nabla}_b\alpha,
\label{eq:xipotential}
\end{equation}
which is equal to the supertranslation $\alpha n^a$ at the boundary
$\Omega=0$. By the considerations after \eqref{eq:confisos}, the
gradient of $\Phi$ is then equivalent to a supertranslation. One notes
that solutions for infinitesimal isometries of flat space
$\partial_a\zeta_b+\partial_b\zeta_a=0$ -- the flat space Killing
equation have a similar decomposition, where
$(\zeta^\mu)_a=\nabla_ax^\mu$ are associated to translations whereas
$(\zeta_{\mu\nu})^a=x_{\mu}(\partial_\nu)^a-x_{\nu}(\partial_\mu)^a$
are the Lorentz generators.

\section{Fluctuations and the general gist}

Before we turn to the specific cases, let us digress over metric
perturbations. Let us suppose that we start with a solution of
Einstein's equation, with possibly a cosmological constant:
\begin{equation}
R_{ab}=\frac{k(D-1)}{\ell^2}g_{ab}.
\label{eq:einsteinspace}
\end{equation} 
One can check \cite{Wald:1984} that, if one changes the metric by a
``small amount'' $\delta g_{ab}=h_{ab}$, the change in the Ricci
tensor is, to first order: 
\begin{equation}
\delta R_{ac}=-\frac{1}{2}\nabla_a\nabla_c {h_b}^b
-\frac{1}{2}\nabla_b\nabla^bh_{ac}
+\nabla_{(c}\nabla^bh_{a)b}+R_{bcad}h^{bd}+{R_{(c}}^dh_{a)d},
\end{equation}
where indices are raised and contracted with the unperturbed metric
$g_{ab}$. The Ricci and Riemann tensor are also computed with respect
to $g_{ab}$. 

The equation for the fluctuations:
\begin{equation}
\delta R_{ab}=\frac{k(D-1)}{\ell^2}\delta g_{ab}
\end{equation}
sets constraints on the $D(D+1)$ dynamical
components of the metric perturbation. However, due to
diffeomorphism invariance, $h_{ab}$ and
$h'_{ab}=h_{ab}+\nabla_a\xi_b+\nabla_b\xi_a$ are physically
indistinguishable for well-defined vector fields $\xi_b$. We can use
this gauge freedom to make $h_{ab}$ traceless $g^{ab}h_{ab}=0$ and
transverse $\nabla^ch_{cb}=0$. Generically, upon a gauge
transformation: 
\begin{equation}
g^{ab}h'_{ab}=g^{ab}h_{ab}+\nabla_c\xi^c,\quad\quad
\nabla^ch'_{cb}=\nabla^ch_{cb}+\nabla^2\xi_b+\nabla^c\nabla_b\xi_c,
\end{equation}
which we will equal to zero to write differential equations for
$\xi_c$. The traceless transverse gauge is achieved by solving:
\begin{equation}
\nabla^b\xi_b=-\thalf g^{ab}h_{ab},\quad\quad
\nabla^2\xi_b+\frac{k(D-1)}{\ell^2}\xi_b=-\nabla^ch_{cb}-\thalf \nabla_bh,
\label{eq:gaugechoice}
\end{equation}
which define the vector field $\xi_c$ up to a solution of the
homogeneous equations:
\begin{equation}
\nabla^b\xi_b=0,\quad\quad
\nabla^2\xi_b+\frac{k(D-1)}{\ell^2}\xi_b=0.
\label{eq:gaugeresidual}
\end{equation}
Both gauge transformations are necessary to reduce the number of
graviton degrees of freedom to their true value -- zero in three
dimensions and two in four. However both solutions assume fast enough
fall-offs at infinity: after all, an acton of the Poincaré group can
change your stationary black hole solution to a moving black hole.

Which brings us to the main point of this paper. The asymptotic
symmetries at the conformal boundary described in the last section
induce, via \eqref{eq:gaugeresidual}, an ``active'' transformation in
the interior. By the discussion above, these are physical: they do
change the physical properties of the background. Specifically, one
can follow \cite{Iyer:1994ys} and associate with an infinitesimal
coordinate transformation $\xi^a$ a charge, for instance, in flat pure
Einstein-Hilbert theory:
\begin{equation}
Q[\xi]=\int_{\Sigma}\epsilon_{aba_1\ldots a_{D-2}}\nabla^a\xi^b,
\end{equation}
one can recover the total mass $M$ associated to time translations and
total angular momentum $J$ associated to azimuthal rotations. The status
of a charge associated to the generic solution of \eqref{eq:confisos}
is less clear. In the Brown-Henneaux case, the Weyl subgroup of the
asymptotic algebra consists of exactly two charges, associated with
$\ell_0$ and $\bar{\ell}_0$, from which one can extract the mass and
angular momentum -- see \eqref{eq:btzmetric} below. In the BMS case
all supertranslations commute, so they may be associated to the
space-time ``hair''. Regardless of their interpretation, the action of
the Brown-Henneaux and BMS generators which do not commute with either
$M$  and $J$ cannot be pure gauge transformations because they change
the value for those observables. In the BMS case, we take a
supertranslation $\xi^a=\alpha(\theta,\phi)n^a$ and find that,
asymptotically: 
\begin{equation}
[\alpha(\theta,\phi)\partial_u,\partial_\phi]=
-(\partial_\phi\alpha)\partial_u,  
\label{eq:changeobs}
\end{equation}
which does not vanish, even at the conformal boundary. By placing
$\xi^a$ as a boundary condition of the residual transformation
\eqref{eq:gaugeresidual}, one can then induce a coordinate
transfomation at the interior which is physical, and may be associated
to a true degree of freedom. In the remaining of the paper we will
illustrate this idea with the BTZ and the Kerr black hole to construct
coordinate transformations at the horizon, as in
\cite{Hawking:2015qqa,Donnay:2015abr}. We use BTZ as a ``proof of
concept'' because the whole structure is readily integrable. The
particular problem of relating asymptotic to near horizon symmetries
in BTZ was tackled by Compère {\it et al.} \cite{Compere:2015knw},
albeit using a dynamic-dependent symplectic form, and proposals for
BTZ hair have been presented by Afshar {\it et al.} in
\cite{Afshar:2016wfy}. The program has also been carried on for the
particular Schwarzschild case by Compère and Long
\cite{Compere:2016jwb,Compere:2016hzt}.

\section{The three-dimensional BTZ black hole}

The BTZ metric for a asymptotically ${\rm AdS}_3$ ($k=-1$) black hole
with mass $M=r_+^2+r_-^2$ and angular momentum $J=2r_+r_-$:  
\cite{Banados:1992wn,Carlip:1995qv},
\begin{equation}
ds^2=-\frac{(r^2-r_+^2)(r^2-r_-^2)}{r^2}dt^2+
\frac{r^2dr^2}{(r^2-r_+^2)(r^2-r_-^2)}
+r^2\left(d\phi-\frac{r_+r_-}{r^2}dt\right)^2
\label{eq:btzmetric}
\end{equation}
is an asymptotically ${\rm AdS_3}$ space-time with $\ell=1$, which can  
be seen heuristically from the large $r$ expansion of
\eqref{eq:btzmetric}. The difference between the BTZ metric and usual
${\rm AdS_3}$ is a global one: upon the change of variables: 
\begin{equation}\label{eq:btzcoordinates}
u=\frac{r_+-r_-}{2}(\phi+t),\quad\quad
\cosh^2\varrho = \frac{r^2-r_-^2}{r_+^2-r_-^2},
\quad\quad v=\frac{r_++r_-}{2}(\phi-t)
\end{equation}
one recovers the left-right invariant metric of the ${\rm SL(2,\mathbb{R})}$
group manifold:
\begin{equation}
ds^2=d\varrho^2+du^2+dv^2+2\cosh2\varrho\, du\,dv.
\label{eq:ads3metric}
\end{equation}

Let us exploit the ${\rm SL(2,\mathbb{R})}$
symmetry of the BTZ geometry. Define the Killing-Cartan form:
\begin{equation}
\eta^{ij}=\tfrac{1}{2}\Tr(\sigma^i\sigma^j)=\begin{pmatrix}
1 & 0 & 0 \\
0 & 0 & \thalf \\
0 & \thalf & 0
\end{pmatrix},
\end{equation}
and its inverse $\eta_{ij}$. From the Euler decomposition of the ${\rm
  SL(2,\mathbb{R})}$ group manifold:
\begin{equation}
g=e^{u\sigma^3}e^{\varrho\sigma^1}e^{v\sigma^3}=
\begin{pmatrix}
e^{u+v}\cosh\varrho & e^{u-v}\sinh\varrho \\
e^{-u+v}\sinh\varrho & e^{-u-v}\cosh\varrho
\end{pmatrix}.
\label{eq:eulerparam}
\end{equation}  
we define the covariant current components:
\begin{equation}
J_i=\thalf\eta_{ij}\Tr(dgg^{-1}\sigma^j),\quad\quad
\bar{J}_i=\thalf\eta_{ij}\Tr(g^{-1}dg\sigma^j).
\end{equation}
Using the inverse metric $g^{ab}$ to \eqref{eq:ads3metric}, we
associate to each component a vector field $J_i^a$ and
$\bar{J}_i^a$. Explicitly:
\begin{equation}
\begin{gathered}
J_3=\frac{\partial}{\partial u},\quad\quad
J_+=e^{2u}\left[\frac{\partial}{\partial \varrho}-\frac{\cosh
    2\varrho}{\sinh 2\varrho}\frac{\partial}{\partial
    u}+\frac{1}{\sinh 2\varrho}\frac{\partial}{\partial v}\right],\\
J_-=e^{-2u}\left[\frac{\partial}{\partial \varrho}+\frac{\cosh
    2\varrho}{\sinh 2\varrho}\frac{\partial}{\partial u}-
 \frac{1}{\sinh 2\varrho}\frac{\partial}{\partial v}\right];
\end{gathered}
\end{equation}
\begin{equation}
\begin{gathered}
\bar{J}_3=\frac{\partial}{\partial v},\quad\quad
\bar{J}_+=e^{-2v}\left[\frac{\partial}{\partial \varrho}-\frac{1}{\sinh
    2\varrho}\frac{\partial}{\partial u}+\frac{\cosh 2\varrho}{\sinh
    2\varrho}\frac{\partial}{\partial v}\right], \\
\bar{J}_-=e^{2v}\left[\frac{\partial}{\partial \varrho}+\frac{1}{\sinh
    2\varrho}\frac{\partial}{\partial u}-\frac{\cosh 2\varrho}{\sinh
    2\varrho}\frac{\partial}{\partial v}\right].
\label{eq:sltworleftright}
\end{gathered}
\end{equation}
Which satisfies the algebra:
\begin{gather}
[J_3,J_\pm]=\pm 2J_\pm,\quad [J_+,J_-]=4 J_3,\quad [J_i,\bar{J}_j]=0\\
[\bar{J}_3,\bar{J}_\pm]=\mp 2 \bar{J}_\pm,\quad [\bar{J}_+,\bar{J}_-]=-4 \bar{J}_3.
\end{gather}
Where commutators are represented by usual vector field brackets (Lie
derivative). From these we define the structure constants
${C_{ij}}^k$, via $[J_i,J_j]={C_{ij}}^kJ_k$ and the
antisymmetric tensor $\epsilon_{ijk}={C_{ij}}^l\eta_{lk}$. One can
show that $\epsilon_{3+-}=4$, $\epsilon^{3+-}=-1$ and that:
\begin{equation}
\epsilon^{ijk}\epsilon_{ijk}=3!\det(\eta_{ij})=-24.
\end{equation}
A little algebra shows that:
\begin{equation}
\nabla_a(J_i)_b=\epsilon_{abc}J_i^c,\quad\quad 
\nabla_a(\bar{J}_i)_b=-\epsilon_{abc}\bar{J}_i^c,
\label{eq:currentderivative}
\end{equation}
where $\epsilon_{abc}$ is the volume form extracted from
\eqref{eq:ads3metric}. 

Following the discussion from last section, we will define ``large
gravitons'' as the solutions of the homogeneous equations
\eqref{eq:gaugeresidual} which assymptote to the Brown-Henneaux
generators at the boundary \eqref{eq:brownhenneaux}. Defining the
scalars $\xi_i=J_i^a\xi_a$, we have, using
\eqref{eq:currentderivative}, that the vector Laplacian can be written
as: 
\begin{equation}
J^a_i\nabla^2\xi_a=\nabla^2\xi_i+{\epsilon_i}^{jk}\nabla_j\xi_k
+2\xi_i,\quad\quad 
\bar{J}^a_i\nabla^2\xi_a=\nabla^2\bar{\xi}_i-{\epsilon_i}^{jk}
\bar{\nabla}_j\bar{\xi}_k+2\xi_i, 
\end{equation}
where $\nabla_i=J_i^a\nabla_a$ -- and analogously for
$\bar{\nabla}_i$, or simply the directional derivative
on the direction $J_i$. The second order equation for $\xi_a$ can now
be cast as a system of first order equations for $\xi_i$:
\begin{equation}
{\epsilon_i}^{jk}\nabla_j\xi_k=\sigma_i,\quad\quad
{\epsilon_i}^{jk}\nabla_j\sigma_k=4\nabla^2\xi_i-4\sigma_i-
4\nabla_i(\nabla_j\xi^j)=-8\sigma_i.
\end{equation}
For the case considered here, we can solve both equations, zero
divergence and zero Laplacian, by setting $\sigma_i=0$. Incidentally,
for the case of ``massive perturbations'', the equation is similar to
a normal mode: 
\begin{equation}
{\epsilon_i}^{jk}\nabla_j\xi_k=\mu\xi_i,
\end{equation}
where $\mu$ is related to the mass of the perturbation, and $\xi_i$ is
now a non-trivial linear combination of the vector field components
and its rotational derivative $\sigma_i$.

The equations for $\bar{\xi}_i=\bar{J}^a_i\xi_a$ are obtained similarly:
\begin{equation}
{\epsilon_i}^{jk}\bar{\nabla}_j\bar{\xi}_k=\bar{\sigma}_i, \quad\quad 
{\epsilon_i}^{jk}\bar{\nabla}_j\bar{\sigma}_k=4\nabla^2\bar{\xi}_i+
4\bar{\sigma}_i-
4\bar{\nabla}_i(\bar{\nabla}_j\bar{\xi}^j)=8\bar{\sigma}_i. 
\end{equation}
For massless modes, one then has:
\begin{equation}
{\epsilon_i}^{jk}\nabla_j\xi_k=
{\epsilon_i}^{jk}\bar{\nabla}_j\bar{\xi}_k=0.
\label{eq:vectorfield}
\end{equation}
One can see that these equations are equivalent by introducing the matrix:
\begin{equation}
L_{ij}=g_{ab}J^a_i\bar{J}^b_j,
\end{equation}
satisfying ${L_i}^j{L^k}_j=\delta^k_i$, with indices lowered and
raised by $\eta$. Vectors -- and tensors -- can be decomposed on
either basis, and ${L_i}^j$ implements the change: 
\begin{equation}
\xi_j={L_i}^j\bar{\xi}_j,\quad\quad
\bar{\xi}_i=\xi_j{L^j}_i.
\end{equation}

The strategy for solving \eqref{eq:vectorfield} for $\xi_a$ which
asymptotes to the Brown-Henneaux generators is simplified due to the
fact that $\ell_n$ only depends on $u$ and $\bar{\ell}_n$ only on
$v$. Since the ``squared'' operator 
\begin{equation}
{\epsilon_i}^{jk}\nabla_j({\epsilon_k}^{lm}\nabla_l\xi_m)=4\nabla^2\xi_i
\end{equation}
is proportional to the scalar Laplacian, each component of $\xi_i$ satisfies:
\begin{equation}
\nabla^2\xi_i=\left[\frac{\partial}{\partial
  z}\left(z(z-1)\frac{\partial}{\partial z}\right)
+\frac{1}{16z(z-1)}\left((2z-1)\frac{\partial^2}{\partial u \partial
    v}-\frac{\partial^2}{\partial u^2}-\frac{\partial^2}{\partial v^2}
\right)\right]\xi_i=0,
\label{eq:laplaciansl2r}
\end{equation}
where $z=\cosh^2\varrho$. Assuming the solution for $\xi_3$ is
only a function of $z$ and $u$ the solution is given readily:
\begin{equation}
\xi_3=-\frac{p^2-1}{2}\left[c_+\left(\frac{z-1}{z}\right)^{p/2}+
  c_-\left(\frac{z}{z-1}\right)^{p/2}\right]e^{2p u}, 
\label{eq:xi3sln}
\end{equation}
and the other components are obtained from
${\epsilon_i}^{jk}\nabla_j\xi_k=0$:
\begin{gather}
\xi_+=\frac{p(p-1)}{2}\left[c_+\left(\frac{z-1}{z}\right)^{(p+1)/2}+
  c_-\left(\frac{z}{z-1}\right)^{(p+1)/2}\right]e^{2(p+1) u},\\
\xi_-=-\frac{p(p+1)}{2}\left[c_+\left(\frac{z-1}{z}\right)^{(p-1)/2}+
c_-\left(\frac{z}{z-1}\right)^{(p-1)/2}\right]e^{2(p-1) u},\\
\end{gather}
where $2p$ is the eigenvalue for $J_3$. By the same token, we have for
$\bar{\xi}_i$ satisfying
${\epsilon_i}^{jk}\bar{\nabla}_j\bar{\xi}_k=0$ only depending on $z$
and $v$:
\begin{gather}
\bar{\xi}_3=\frac{q^2-1}{2}\left[\bar{c}_+\left(\frac{z-1}{z}\right)^{q/2}+
  \bar{c}_-\left(\frac{z}{z-1}
  \right)^{q/2}\right]e^{2q v},\\
\bar{\xi}_+=\frac{q(q+1)}{2}\left[\bar{c}_+\left(\frac{z-1}{z}\right)^{(q-1)/2}+
  \bar{c}_-\left(\frac{z}{z-1}
  \right)^{(q-1)/2}\right]e^{2(q-1) v},\\
\bar{\xi}_-=-\frac{q(q-1)}{2}\left[\bar{c}_+\left(\frac{z-1}{z}\right)^{(q+1)/2}+
  \bar{c}_-\left(\frac{z}{z-1} 
  \right)^{(q+1)/2}\right]e^{2(q+1) v},
\end{gather}
The constants sitting in front of the expressions for $\xi_3$ and
$\bar\xi_3$ were chosen so that $\xi^a$ and $\bar\xi^a$ asymptote to
the Brown-Henneaux generators in a simpler expression. 

As one can see, no mention to the actual BTZ metric
\eqref{eq:btzmetric} was made, the whole construction being {\it a
  priori} defined on a sort of covering space where $u$ and $v$ cover
the whole plane. Coming back to the $t$ and $\phi$ coordinates, we
find that $p$ and $q$ are related to the frequency and angular
momentum -- eigenvalues of $\partial_t$ and $\partial_\phi$ by:
\begin{equation}
p=-\frac{i}{2}\frac{\omega-m}{r_+-r_-},\quad\quad
q=\frac{i}{2}\frac{\omega+m}{r_++r_-}.
\end{equation}

Using \eqref{eq:currentderivative} one may
calculate the metric perturbation due to $\xi^a$: $h_{ab} =
\nabla_a\xi_b + \nabla_b\xi_a$. Defining null coordinates on the $\varrho-u$ plane,
$u_{\pm} = u \pm \thalf\log(\tanh\varrho)$, we have:  
\begin{equation}\label{eq:leftgraviton}
	h_{ab} = -2p(p^2-1) \left[ c_+e^{2pu_+}(du_+)_a(du_+)_b +
          c_-e^{2pu_-}(du_-)_a(du_-)_b \right]. 
\end{equation}
Thus, for purely ingoing solution near the black hole horizon
$(\varrho=0)$, we must set $c_-=0$. Note that for a single valued
solution, 
\begin{equation}
	p = i\frac{m}{r_+ - r_-},
\end{equation}
with $m$ integer, in order that $2pu = im(\phi + t)$. So
\eqref{eq:leftgraviton} corresponds to ingoing and outgoing 
waves of frequency equal to minus the angular momentum. The $(\phi + t)$
dependence of the phase indicate the ``chiral'', left-moving character
of these modes. The right-moving modes are calculated from
$\bar\xi_a$. In fact, we need only change $(u,p,c_\pm)\rightarrow
(v,q,\bar c_\pm)$ in 
\eqref{eq:leftgraviton} to obtain the solution: 
\begin{equation}\label{eq:rightgraviton} 
	\bar h_{ab} = -2q(q^2-1) \left[ \bar
          c_+e^{2qv_+}(dv_+)_a(dv_+)_b + \bar
          c_-e^{2qv_-}(dv_-)_a(dv_-)_b \right], 
\end{equation}
where $v_\pm = v\pm\thalf\log(\tanh\varrho)$. The interpretation of
the null coordinates $v_\pm$ are different now: one can check that
they correspond to outgoing and ingoing coordinates, respectively, in 
BTZ. Thus our physical solution requires $\bar c_+ = 0$. According to
\eqref{eq:btzcoordinates}, 
\begin{equation}
	q = i\frac{\bar m}{r_+ + r_-},
\end{equation}
with $\bar m$ integer, in such a way that $2qv = i\bar m(\phi-t)$. 

In terms of the $\varrho,u,v$ coordinates, the diffeomorphism $\xi^a$
can be written as:
\begin{equation}\begin{split}
	\xi^a =
        \,&c_+e^{2pu_+}\left[-\frac{p(p+\cosh2\varrho)}{2\sinh2\varrho}
\,\partial_\varrho 
          +
          \frac{p^2+p\cosh2\varrho+\sinh^22\varrho}{2\sinh^22\varrho}
\,\partial_u
          -
          \frac{p(p\cosh2\varrho+1)}{2\sinh^22\varrho}
\,\partial_v\right]
        \\ 
		+\,&c_-e^{2pu_-}\left[\frac{p(p-\cosh2\varrho)}{2\sinh2\varrho}
\,\partial_\varrho +
\frac{p^2-p\cosh2\varrho+\sinh^22\varrho}{2\sinh^22\varrho}
\,\partial_u -
\frac{p(p\cosh2\varrho-1)}{2\sinh^22\varrho}\,\partial_v\right].  
\end{split}\end{equation} 
At spatial infinity ($\varrho=\infty$), where $u_+ = u_- \rightarrow u$,
\begin{equation}
	\xi^a \rightarrow
        (c_++c_-)e^{2pu}\left(-\frac{p}{2}\frac{\partial}{\partial\varrho} +
          \half\frac{\partial}{\partial u}\right), 
\end{equation}
as expected, since we constructed the diffeomorphism to be the
Brown-Henneaux generators \eqref{eq:brownhenneaux} there. Let us
define two new basis of vectors in which the diffeomorphisms can be
better represented. First, define the coordinate 
\begin{equation}
	\chi = u+v = r_+\left(\phi-\frac{r_-}{r_+}t\right) = r_+\phi_H
\end{equation}
related to the co-rotating angular coordinate $\phi_H$ at the event
horizon. Now, we introduce the basis of vectors $\{\partial_+$,
$\bar\partial_+$, $\partial_{\chi^+}\}$ dual to $du_+$, $dv_+$ and
$d\chi$:   
\begin{equation}
	\partial_+ = \frac{\partial_u - \partial_v}{2} +
        \frac{\sinh2\varrho}{2}\,\partial_\varrho,\quad
	\bar\partial_+ = \frac{\partial_v - \partial_u}{2} +
        \frac{\sinh2\varrho}{2}\,\partial_\varrho,\quad
	\partial_{\chi^+} = \frac{\partial_u + \partial_v}{2} -
        \frac{\sinh2\varrho}{2}\,\partial_\varrho,
\end{equation}
and the corresponding dual basis to $du_-$, $dv_-$ and $d\chi$: 
\begin{gather}
	\partial_- = \frac{\partial_u - \partial_v}{2} -
        \frac{\sinh2\varrho}{2}\,\partial_\varrho,\quad
	\bar\partial_- = \frac{\partial_v - \partial_u}{2} -
        \frac{\sinh2\varrho}{2}\,\partial_\varrho,\quad
	\partial_{\chi^-} = \frac{\partial_u + \partial_v}{2} +
        \frac{\sinh2\varrho}{2}\,\partial_\varrho.
\end{gather}
In these two sets of basis, the diffeomorphisms can be written as
\begin{equation}\begin{split}\label{eq:leftdiffeo}
	\xi^a = \,&c_+e^{2pu_+}\left[\half\,\partial_+ -
          \frac{p(p+1)}{4\sinh^2\varrho}\,\bar\partial_+ +
          \left(\half -
            \frac{p(p-1)}{4\cosh^2\varrho}\right)\,\partial_{\chi^+}\right]\\ 
		+\,&c_-e^{2pu_-}\left[\half\,\partial_- -
                  \frac{p(p-1)}{4\sinh^2\varrho}\,\bar\partial_- +
                  \left(\half -
                    \frac{p(p+1)}{4\cosh^2\varrho}\right)\,\partial_{\chi^-}\right], 
\end{split}\end{equation}
and
\begin{equation}\begin{split}\label{eq:rightdiffeo}
	\bar\xi^a = \,&\bar c_+e^{2qv_+}\left[\half\,\bar\partial_+
          - \frac{q(q+1)}{4\sinh^2\varrho}\,\partial_+ + \left(\half
            -
            \frac{q(q-1)}{4\cosh^2\varrho}\right)\,\partial_{\chi^+}\right]\\ 
		+\,&\bar c_-e^{2qv_-}\left[\half\,\bar\partial_- -
                  \frac{q(q-1)}{4\sinh^2\varrho}\,\partial_- +
                  \left(\half -
                    \frac{q(q+1)}{4\cosh^2\varrho}\right)\,\partial_{\chi^-}\right]. 
\end{split}\end{equation}

Written in terms of the coordinates $u_\pm$ and $v_\pm$, we can better
understand the role of the set of constants $c_\pm$ and
$\bar{c}_\pm$. We begin by related to the usual time
and radial BTZ coordinates $t$ and $r$ in \eqref{eq:btzmetric}, can be
seen to satisfy the asymptotic values 
\begin{equation}\begin{gathered}
	t = \infty: \left\{
	\begin{aligned}
		& u_++u_- = \infty\\
		& v_++v_- = -\infty
	\end{aligned} \right. ,\qquad
	t = -\infty: \left\{
	\begin{aligned}
		& u_++u_- = -\infty\\
		& v_++v_- = \infty
	\end{aligned} \right. ,\\
	r = r_+: \left\{
	\begin{aligned}
		& u_+-u_- = -\infty\\
		& v_+-v_- = -\infty
	\end{aligned} \right. .
\end{gathered}\end{equation}
Hence one can identify the future event horizon $\mathcal{H}^+$ as
located at the outgoing coordinates $u_- = \infty$ and $v_+=-\infty$,
while the past horizon $\mathcal{H}^-$ at the ingoing coordinates
$u_+=-\infty$ and $v_-=\infty$. Now, the left diffeomorphism
\eqref{eq:leftdiffeo} act at the horizons as 
\begin{gather}
	\mathcal{H}^+: \left\{
	\begin{aligned}
		&u_+ \rightarrow u_+ + \frac{c_+}{2}e^{2pu_+}\\
		&\chi \rightarrow \chi + \frac{c_+}{2}\left(1 -
                  \frac{p(p-1)}{2}\right)e^{2pu_+} 
	\end{aligned} \right. ,\\
	\mathcal{H}^-: \left\{
	\begin{aligned}
		&u_- \rightarrow u_- + \frac{c_-}{2} e^{2pu_-}\\
		&\chi \rightarrow \chi + \frac{c_-}{2}\left(1 -
                  \frac{p(p+1)}{2}\right)e^{2pu_-} 
	\end{aligned} \right. ,
\end{gather}
and the right diffeomorphism \eqref{eq:rightdiffeo} as
\begin{gather}
	\mathcal{H}^+: \left\{
	\begin{aligned}
		&v_- \rightarrow v_- + \frac{\bar c_-}{2}e^{2qv_-}\\
		&\chi \rightarrow \chi + \frac{\bar c_-}{2}\left(1 -
                  \frac{q(q+1)}{2}\right)e^{2qv_-} 
	\end{aligned} \right. ,\\
	\mathcal{H}^-: \left\{
	\begin{aligned}
		&v_+ \rightarrow v_+ + \frac{\bar c_+}{2} e^{2qv_+}\\
		&\chi \rightarrow \chi + \frac{\bar c_+}{2}\left(1 -
                  \frac{q(q-1)}{2}\right)e^{2qv_+} 
	\end{aligned} \right. .
\end{gather}
This rather simple form can be cast in a mode-decomposition
independent form, for instance, if $c_-=0$, the transformation of
$\mathcal{H}^+$ is given by:
\begin{equation}
u_+\rightarrow u_++\epsilon f(u_+),\quad\quad
\chi\rightarrow \chi+\epsilon
(f(u_+)+\tfrac{1}{4}f'(u_+)-\tfrac{1}{8}f''(u_+)),
\label{eq:diffeobtz}
\end{equation}
which can be compared with the literature as a different gauge choice
for the large diffeomorphisms. One can also expect this behavior from
the generic solution presented in \cite{Banados:1998gg}. We also note
that the condition $c_-=0$ leaves the past horizon invariant. One can
then, relate the solutions obtained for $c_+=0$ and $c_-=0$ in
\eqref{eq:xi3sln} by a time reversal. A similar condition will arise
in the next section.

\section{The Kerr black hole}

The Kerr metric for mass $M$ and angular momentum $J=aM$ is suitably
described in the Kinnersley null tetrad basis, see
\cite{Chandrasekhar:1985kt}:
\begin{equation}
\begin{gathered}
\ell=e_1=\frac{r^2+a^2}{\Delta}\frac{\partial}{\partial
  t}+\frac{a}{\Delta}\frac{\partial}{\partial \phi}+\frac{\partial}{\partial r},\quad
n=e_2=\frac{\Delta}{2\Sigma}
\left(\frac{r^2+a^2}{\Delta}\frac{\partial}{\partial
  t}+\frac{a}{\Delta}\frac{\partial}{\partial
  \phi}-\frac{\partial}{\partial r}\right) \\
m = e_3= \frac{1}{\sqrt{2}\sigma}\left(
  ia\sin\theta\frac{\partial}{\partial t}+\frac{\partial}{\partial
    \theta}+\frac{i}{\sin\theta}\frac{\partial}{\partial\phi}\right),\quad
\bar{m}=e_4 = (m)^*.
\end{gathered}
\end{equation}
where $t,r,\theta,\phi$ are the Boyer-Lindquist coordinates and 
\begin{equation}
\Delta=r^2-2Mr+a^2=(r-r_+)(r-r_-),\quad
\sigma=r+ia\sin\theta,\quad \Sigma=\sigma\sigma^*=|\sigma|^2,
\end{equation}
and the metric is defined by saying that the only non-vanishing inner
products are:
\begin{equation}
\ell\cdot n=-1,\quad\quad m\cdot\bar{m}=1.
\end{equation}
This defines the local Minkowski metric to be:
\begin{equation}
\eta^{\mu\nu}=\begin{pmatrix}
0 & -1 & 0 & 0 \\
-1 & 0 & 0 & 0 \\
0 & 0 & 0 & 1 \\
0 & 0 & 1 & 0 \\
\end{pmatrix}.
\end{equation}
The Kinnersley basis is particularly useful for dealing with
perturbations of the Kerr metric. Separability of the equations is
achieved for scalar, spinorial, vector and gravitational
perturbations. Our problem is to find non-trivial solutions to the
equations
\begin{equation}
\nabla^2\xi_a=0,\quad\quad \nabla_a\xi^a=0,
\label{eq:kerr-perturbations}
\end{equation}
which approach supertranslations at null infinity ${\cal I}^\pm$. In
general, the problem can be cast into the solution for the potential
of a spin-1 perturbation of the Kerr black hole. These are defined by
the solution to the vacuum Maxwell equations
\begin{equation}
\nabla^{a}F_{ab}=\nabla_{[a}F_{bc]}=0
\label{eq:maxwell}
\end{equation}
in terms of the vector potential $A_a$ such that
$F_{ab}=\nabla_aA_b-\nabla_bA_a$. We have that the vacuum Maxwell
equations generate a solution to \eqref{eq:kerr-perturbations} if we
choose the Lorenz gauge $\nabla_aA^a=0$, given that the Kerr metric
is Ricci-flat. Expressions for the general potential in
\cite{Chrzanowski:1975wv} use the ``ingoing and outgoing radiation
gauges'' $\ell^aA_a=0$ or $n^aA_a=0$, which do not suit our purposes
because the gauge transformation is as hard as the problem we want to
solve. As in \eqref{eq:xipotential}, we will suppose that $\xi_a$ is a
gradient $\xi_a=\nabla_a\Phi$ to begin with. Then one can obtain a
solution to \eqref{eq:kerr-perturbations} by just considering $\Phi$
to be a solution of the wave equation $\nabla^2\Phi=0$, with the
asymptotic boundary condition from \eqref{eq:xipotential}:
\begin{equation}
\Phi\approx \Omega \alpha(\theta,\phi),\quad\text{at}\quad\Omega=0.
\label{eq:xipotentialbc}
\end{equation} 
Again, the divergence-free condition is satisfied by assumption and:
\begin{equation}
\nabla^2\nabla_b\Phi=\nabla_b\nabla^2\Phi+{R_{b}}^d\nabla_d\Phi
\end{equation}
also vanishes in Ricci-flat backgrounds like the the Kerr metric. 
The scalar Laplacian can be written in terms of the Ricci rotation
coefficients
$\gamma_{\rho\mu\nu}=(e_\rho)^a(e_\mu)^b\nabla_a(e_\nu)_b$:
\begin{equation}
\nabla^2\Phi=\eta^{\mu\nu}(e_\mu)^a(e_\nu)^b\nabla_a\nabla_b\Phi
=\eta^{\mu\nu}e_{\mu}(e_{\nu}(\Phi))-
\eta^{\mu\nu}\gamma_{\rho\mu\nu}\eta^{\rho\sigma}e_{\sigma}(\Phi), 
\end{equation}
where we used the definition of vectors as directional derivatives to
write $(e_\mu)^a\nabla_a\Phi\equiv e_\mu(\Phi)$. In terms of the
Newman-Penrose symbols, which for the Kerr metric can be found in 
\cite{Chandrasekhar:1985kt}, the Laplacian becomes:  
\begin{multline}
\nabla^2\Phi=
-\ell(n(\Phi))-n(\ell(\Phi))+m(\bar{m}(\Phi))+\bar{m}(m(\Phi))\\
+(\bar{\rho}+\bar{\rho}^*-\bar{\epsilon}-\bar{\epsilon}^*)n(\Phi)
+(\bar{\gamma}^*+\bar{\gamma}-\bar{\mu}^*-\bar{\mu})\ell(\Phi)\\
-(\bar{\tau}^*-\bar{\pi}+\bar{\alpha}-\bar{\beta}^*)m(\Phi)
-(\bar{\tau}-\bar{\pi}^*+\bar{\alpha}^*-\bar{\beta})\bar{m}(\Phi).
\end{multline}
We continue by introducing the differential operators:
\begin{equation}
\begin{gathered}
\ell = D =\frac{r^2+a^2}{\Delta}\frac{\partial}{\partial
  t}+\frac{a}{\Delta}
\frac{\partial}{\partial\phi}+\frac{\partial}{\partial r},\quad
n=-\frac{\Delta}{2\Sigma}\bar{D},\\
m=\frac{1}{\sqrt{2}\sigma}Q=\frac{1}{\sqrt{2}\sigma} \left(
ia\sin\theta\frac{\partial}{\partial t}+\frac{i}{\sin\theta}
\frac{\partial}{\partial\phi}+\frac{\partial}{\partial\theta}\right),\quad
\bar{m}=\frac{1}{\sqrt{2}\sigma^*}\bar{Q},
\end{gathered}
\end{equation}
in terms of which we can write the Laplacian operator as a sum of two
anticommutators: 
\begin{equation}
\nabla^2\Phi= \frac{1}{\Sigma}\left(\frac{1}{2\Delta}\{\Delta D,\Delta
  \bar{D}\} +\frac{1}{2\sin^2\theta}\{\sin\theta Q,\sin\theta\bar{Q}\}\right).
\end{equation}
It can be checked that both terms of the sum in the brackets
commute. Writing each explicitly in terms of the Boyer-Lindquist
coordinate operators:
\begin{gather}
\frac{1}{2\Delta}\{\Delta D,\Delta \bar{D}\}
=\frac{\partial}{\partial r}\Delta\frac{\partial}{\partial r}
-\frac{1}{\Delta}\left((r^2+a^2)\frac{\partial}{\partial
    t}+a\frac{\partial}{\partial\phi}\right)^2,\\
\frac{1}{2\sin^2\theta}\{\sin\theta Q,\sin\theta\bar{Q}\}=
\frac{1}{\sin\theta}\frac{\partial}{\partial
  \theta}\sin\theta\frac{\partial}{\partial\theta}
+\frac{1}{\sin^2\theta}\left(a\sin^2\theta\frac{\partial}{\partial
    t}+\frac{\partial}{\partial\phi} \right)^2.
\end{gather}
 The wave operator is now separable and the solutions can be written
 in terms of confluent Heun equations. We will write the generic
 solution for frequency $\omega$, angular momentum $m$ and angular
 quantum number $l$ by: 
\begin{equation}
\Phi_{\omega,l,m}(r,\theta\phi)= e^{-i\omega t}(
C^\infty_+ h^+_{\omega,\ell,m}(r)+C^\infty_- 
h^-_{\omega,\ell,m}(r))\,{_0S_{lm}}(\cos\theta) e^{im\phi},
\label{eq:scalarsolution}
\end{equation}
where $h^\pm$ are confluent Heun functions \cite{ronveaux1995heun}
and ${_0S_{lm}}$ are the scalar spheroidal harmonics
\cite{Berti:2005gp}. In the zero frequency limit, $l$ 
reduces to the usual spherical quantum number, ${_0S_{lm}}$ to the
usual spherical harmonic $Y_l^m$ and the separation constant between
the angular and radial equation is $l(l+1)$. The radial functions have
the asymptotic behavior -- see \cite{daCunha:2015ana}: 
\begin{equation}
h^\pm_{\omega,l,m}=\frac{e^{\pm i\omega
    r}}{r^{1\mp i\omega (r_++r_-)}}(1+{\cal O}(r^{-1})),
\label{eq:asymptradial}
\end{equation}
which poses us a problem for the boundary conditions at ${\cal
  I}^\pm$, given by \eqref{eq:xipotentialbc}, because
\eqref{eq:asymptradial} has an essential singularity at
$\Omega\simeq (r\pm t)^{-1}=0$, unless $\omega=0$. 

One can actually show that there is a solution: if one goes to the
unphysical metric $g_{ab}$ a little algebra shows that if $\Phi$
satisfies the wave equation with respect to the metric $g_{ab}$, then
$\hat{\Phi}=\Omega^{-1}\Phi$ satisfies:
\begin{equation}
\hat{g}^{ab}\hat{\nabla}_a\hat{\nabla}_b
\hat{\Phi}-\tfrac{1}{6}\hat{R}\hat{\Phi}=0, 
\label{eq:confpot}
\end{equation}
where $\hat{R}$ is the Ricci scalar associated with $\hat{g}_{ab}$. As
$\Omega=0$, the asymptotic flat space-time approaches the asymptotic
structure of Minkowski space, and there is a gauge choice where
$\hat{g}_{ab}$ approaches the metric of the standard Einstein static
universe:
\begin{equation}
d\hat{s}_{\rm EE}^2=-dx^2+dy^2+\sin^2y(d\theta^2+\sin^2\theta\,d\phi^2),
\end{equation}
which has constant Ricci scalar: $\hat{R}_{\rm EE}=24$. In this case,
one can verify \cite{BirrellDavies} that \eqref{eq:confpot} has a
Green's function:
\begin{equation}
G_{\rm EE}=\frac{1}{2\pi^2}\frac{1}{\cos(\Delta x)-\cos(\Delta s)},
\end{equation}
where 
\begin{equation}
\Delta s=\cos y\cos y'+\sin y\sin y'(\cos\theta\cos\theta'+
\sin\theta\sin\theta'\cos(\phi-\phi'))
\end{equation}
is the invariant distance on the $3$-sphere. One sees that ${\cal
  I}^\pm$ is mapped to $x\pm y=\pi$, where the static universe metric
can be continued without any problems. By propagating using the left
or right-moving part of $G_{\rm EE}$ one can obtain a solution at the
interior with the prescribed boundary behavior. Note that
$\hat{\Phi}=\alpha$ is finite at $\Omega=0$.

As in the case with the BTZ black hole, there is an ambiguity with the
choice of constants $c_\pm^\infty$.  Like there, we may fix this by
requiring that the induced diffeomorphism is ``outward moving'' near
the outer horizon $r_+$. As it happens, for $\omega=0$ the radial
equation simplifies to Riemann's differential equation form, and the
generic solution can be written in terms of Gauss' hypergeometric
functions. Near the outer horizon, one can find the asymptotic
behavior:
\begin{equation}
\Phi_{l,m}(r)\approx (c^+_+(r-r_+)^{i\theta_+} +c^+_-(r-r_-)^{-i\theta_+}
)P_l^m(\cos\theta)e^{im\phi},\quad\text{near}\quad r=r_+. 
\end{equation}
However, from the discussion at \eqref{eq:changeobs}, if one takes the
viewpoint that the ``only true observables'' are the mass and the
angular momentum, the dependence on $\theta$ above is spurious. One
then is led to consideration of the asymptotic potentials of the form:
\begin{equation}
\Phi_{l,m}(r)\approx c^+_+(r-r_+)^{i\theta_+}e^{im\phi}+
c^+_-(r-r_-)^{-i\theta_+}e^{im\phi},\quad\text{at}\quad r=r_+.
\label{eq:phyasymptrp} 
\end{equation}
The coefficients at the horizon $c^+_\pm$ are linear combinations of 
$c^\infty_\pm$, the linear transformation matrix entries are called
the {\it connection coefficients}. The radial exponent has the nice
interpretation: 
\begin{equation}
\theta_+=\frac{r_-
  m(r_++r_-)}{r_+-r_-}=\frac{1}{2\pi}\frac{m\Omega_+}{T_+},
\end{equation}
where $\Omega_+$ and $T_+$ are the angular velocity and the
temperature of the horizon at $r=r_+$. Therefore, $\theta_+$ is, up to
a factor of $2\pi$, the
increase in entropy of the black hole by absorption of a wave with
zero energy and angular momentum $m$.

In terms of the tortoise radial coordinate:
\begin{equation}
r_*=r+\frac{r_+(r_++r_-)}{r_+-r_-}\log\left(\frac{r}{r_+}-1\right)
-\frac{r_-(r_++r_-)}{r_+-r_-}\log\left(\frac{r}{r_-}-1\right),
\end{equation}
the coordinate transformation can be interpreted as a chiral
shift of the co-moving coordinates and a shift of $r$, which places a
conformal transformation. The transformation is best seen in
Teukolsky-Eddington-Finkelstein coordinates \cite{Teukolsky:1973jk} --
see also \cite{Misner:1974qy}. Defining $\tilde{\phi}_\pm$ as 
\begin{equation}
d\tilde{\phi}_\pm = d\phi \pm \frac{a}{r^2+a^2}dr_*=
d\phi \pm \frac{a}{\Delta}dr,
\end{equation}
we can put the covariant form of the transformation in the rather
simple form:
\begin{equation}
(\xi_m)_a=im\, c^+_+ e^{im\tilde{\phi}_+}(d\tilde{\phi}_+)_a
+im\, c^+_- e^{im\tilde{\phi}_-}(d\tilde{\phi}_-)_a
\end{equation}
The structure is then quite similar to \eqref{eq:leftgraviton} and
\eqref{eq:rightgraviton}. In terms of vector fields (supposing either
$c^+_\pm=0$, and $r\simeq r_+$): 
\begin{equation}
(\xi_m)^a\propto im\, e^{im\tilde{\phi}_\pm}\frac{1}{\Sigma \Delta}\left(
\frac{\partial}{\partial t}+\Omega_+\frac{\partial}{\partial \phi}
\pm \frac{\partial}{\partial r_*}\right),
\end{equation}
approaching the ``radial-temporal'' $\ell,n$ elements of Kinnersley
null tetrad near the outer horizon, depending on whether one chooses
purely ingoing or outgoing transformations.

\begin{figure}[htb]
\begin{center}
\mbox{\includegraphics[width=0.4\textwidth]{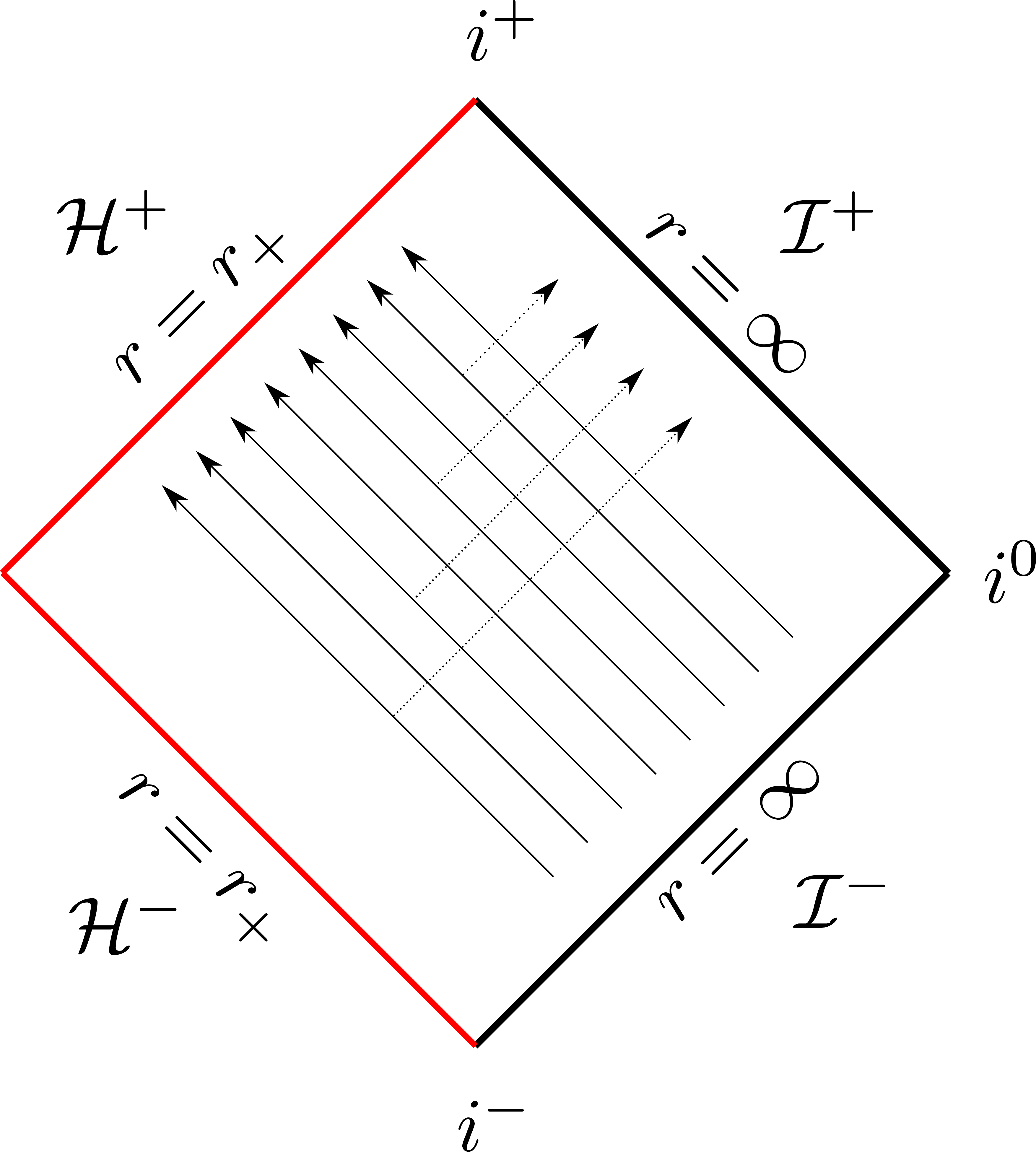}}
\hspace{10mm}
\mbox{\includegraphics[width=0.4\textwidth]{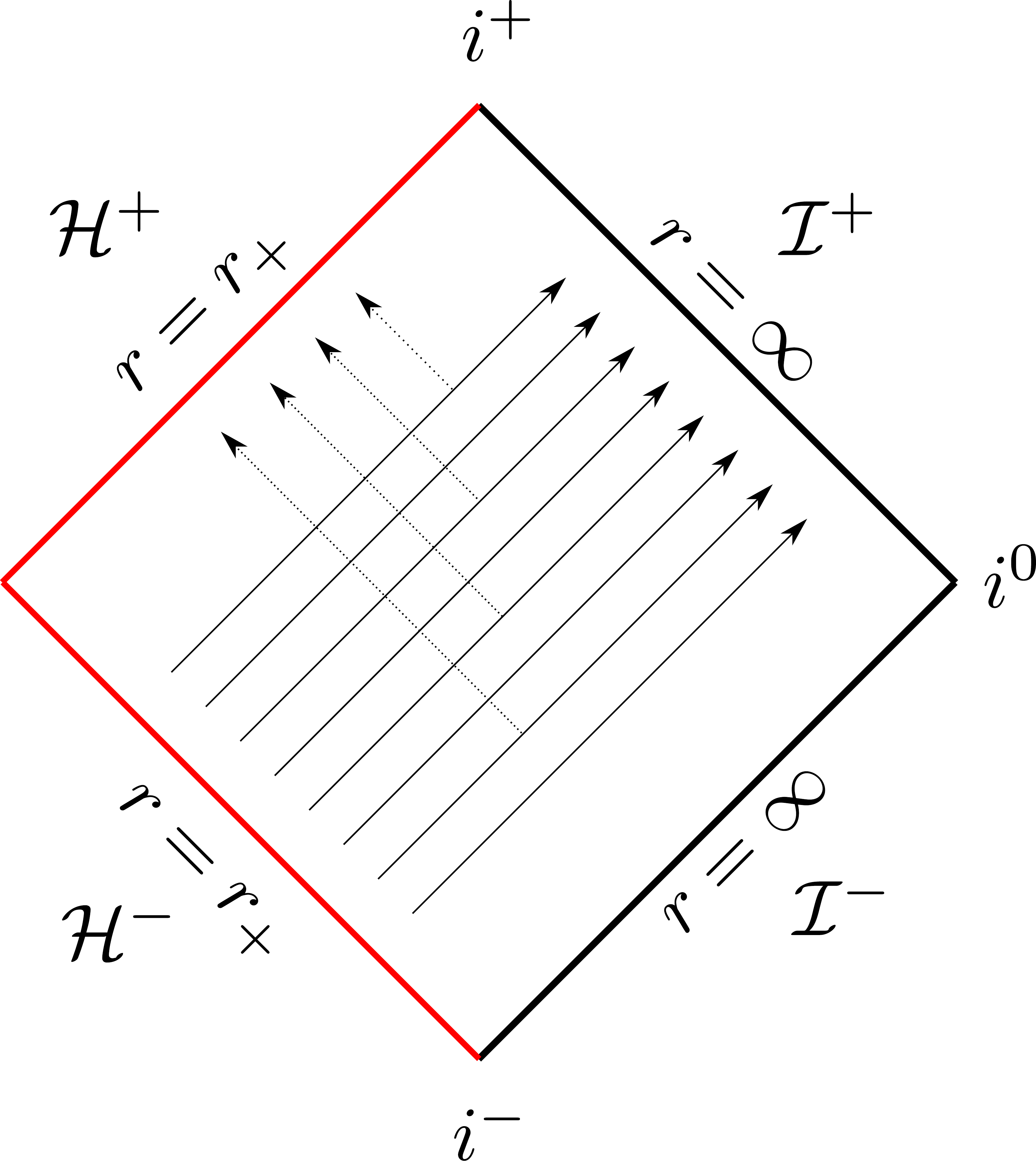}}
\caption{Respectively, the in- and up-mode of wave scattering at the
  outer horizon of a black hole. The in-mode is characterized by the
  solution of the wave equation in the Kerr background with
  $c^+_-=0$ whereas in the up-mode $c^\infty_-=0$ is chosen.}
\label{fig:in-up}
\end{center}
\end{figure}

However, in taking purely ingoing modes at the
horizon, one has to fix $c^\infty_+$ and $c^\infty_-$
appropriately. This fixing seems strange from the boundary conditions 
placed at ${\cal I}^+$, since this condition is actually placed at
${\cal H}^+$. The situation resembles to the ``in-mode''
vs. ``up-mode'' decomposition of black hole scattering, see
Fig. \ref{fig:in-up}. In terms of the actual choice of solution of the
wave equation both of them use the time reversal symmetry to identify
the ``outgoing'' function at the future null infinity ${\cal I}^+$ or
future horizon ${\cal H}^+$ with the ``ingoing'' function at past null
infinity ${\cal I}^-$ or the past horizon ${\cal   H}^-$. The use of
time reversal is necessary because, generically, there is no canonical
way to associate the space of physical states at ${\cal I}^+$ to the
one at ${\cal I}^-$, and hence no way to compute scattering
coefficients. By the time reversal recipe, the coefficients of
normalized waves have the interpretation of scattering coefficients.  

For our purposes, the in-mode choice selects the unique solution by
setting $c^+_-=0$. One then faces the problem of interpreting the
$u^\infty_-$ component in \eqref{eq:scalarsolution}. We will interpret
this just as the scattering problem: as a time-reversal induced
supertranslation at ${\cal I}^-$. By this recipe, the outgoing
supertranslations associated to the horizon are very specific
combinations of supertranslations at ${\cal I}^+$ and ${\cal
  I}^-$. From a physical point of view, this interpretation seems in
league with the experiments performed by an asymptotic observer in
order to detect the black hole. We note that this particular feature
distinguishes the Kerr case from the BTZ case, with the former
believed to be closer to the generic case.

\section{Discussion}

In this paper we proposed to use the gauge fixing procedure for
perturbations in classical general relativity to associate asymptotic
symmetries to bulk symmetries. In the case where the space-time has a
black hole, these will induce transformations at the horizon, in a way
reminiscent of recent proposals by
\cite{Hawking:2015qqa,Averin:2016ybl,Donnay:2015abr,Compere:2015knw,
  Donnay:2016ejv}
-- among others. One finds that these have always ``zero-frequency'',
corresponding to ``soft-gravitons'' in the IR-limit, which have been
discussed by numerous authors -- see
\cite{Balachandran:2016ohv,Balachandran:2016bqj} for a recent
overview. The induced transformations at the horizon seem to have the
generic behavior of coordinate transformations involving the
co-rotating angular variable $\phi_R$, along with 
suitable scalings on the radial direction. One also sees that both in
three and four dimensions the condition that these transformations are
outgoing at the horizon ${\cal H}^+$ selects the solution
uniquely. Because of the non-commuting nature of (some of) the
supertranslations at infinity to the mass and angular momentum, we know
that these constitute ``large gauge transformations'' and should be
treated as true degrees of freedom of the theory.

Given the generality of the elements, one cannot help but wonder
whether the construction generalizes to higher dimensions. In a series of
articles, Barnich and collaborators
\cite{Barnich:2009se,Barnich:2010eb,Barnich:2011mi,Barnich:2013axa}
helped with the issue of supertranslations in different dimensions and
settings, so it seems an easy target. The connections of the
diffeomorphisms outlined here and conformal field theory
\cite{Carlip:1999cy,Barnich:2012rz} cannot be overlooked. The existence
of constraints posed by the gauge choice \eqref{eq:gaugeresidual} are
the key piece behind the appearance of central terms 
in the representation of the conserved quantities in classical
mechanics \cite{Arnold:1989}. A positive
result for the appearance of central terms in the representation
algebra would be most interesting in the long standing problem of
holographic description of  asymptotically flat space-times
\cite{Arcioni:2003td,Arcioni:2003xx}. We also hope to address the
problem of quantization in future work.

\section*{Acknowledgements}

The authors thank Amílcar de Queiroz, Francisco Brito and
A. P. Balachandran for discussions and suggestions. 
BCC acknowledges partial support from PROPESQ/UFPE and FACEPE
for support under grant no. APQ-0051-1.05/15. FR acknowledges
financial support from CNPq - Brazil. 


\providecommand{\href}[2]{#2}\begingroup\raggedright\endgroup

\end{document}